\documentclass[article]{aa}
\usepackage{graphicx}

\def\figfilter{
\begin{figure*}
\resizebox{\hsize}{!}{\includegraphics{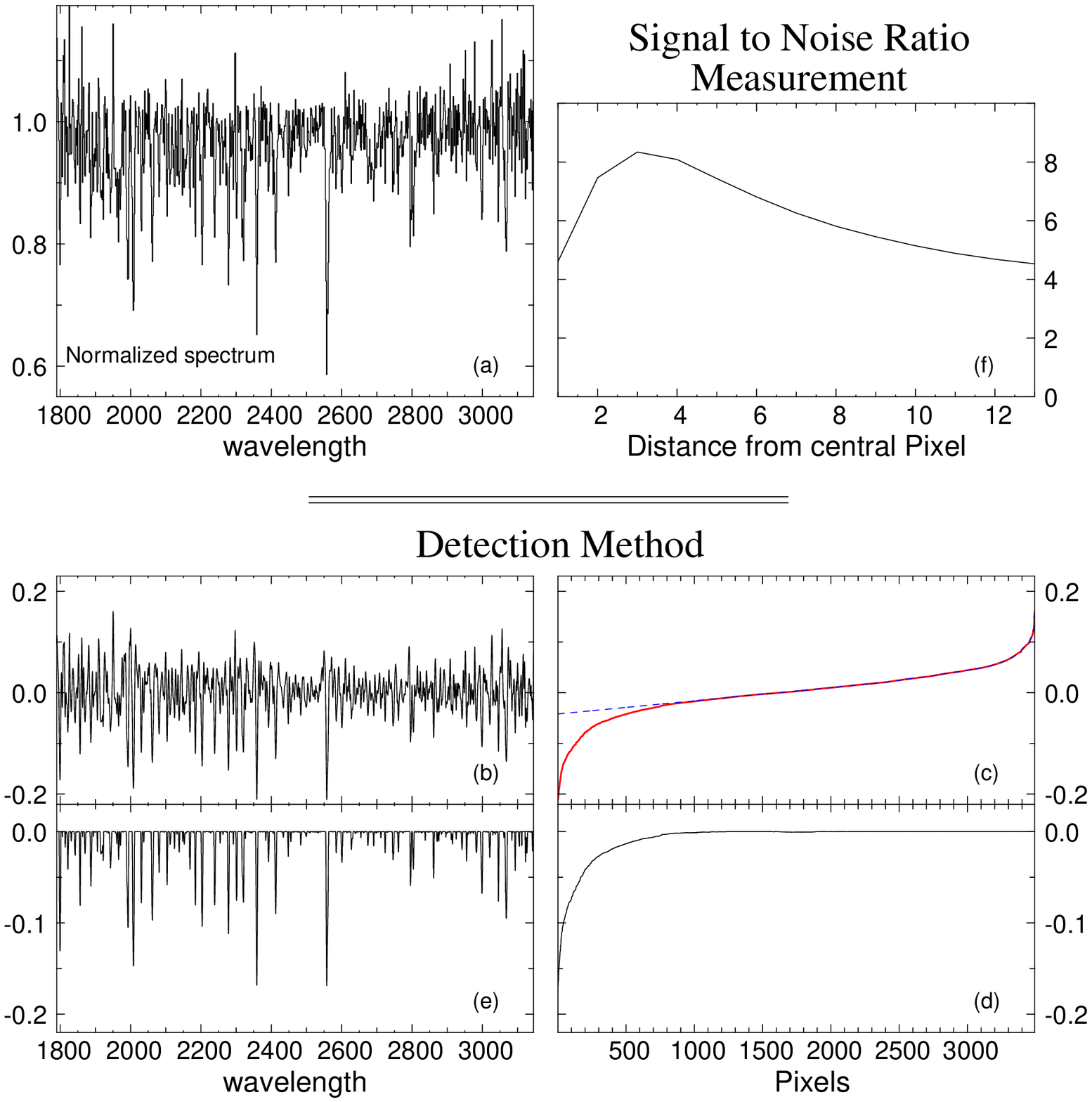}}
\caption{ {  Illustration of the  procedure used to  detect absorption
lines.  Panel (a) represents the original normalized spectrum which is
wavelet  filtered to  select scales  characteristic of  the absorption
lines. The filtered spectrum is shown in panel b. All pixels in  the spectrum are  sorted according to
their value (panel  c).  Low values correspond to  absorption and high
values to spikes; the intermediate values are dominated by noise which
is fitted  by the dashed  line in panel  (c).  This level of  noise is
subtracted to  each pixel to give  the new distribution  of panel (d)
and pixels  are reordered.  The resulting spectrum  is shown  in panel
(e).   This spectrum  is  used only  to select  the regions  where
absorption  lines   are  detected.   The  equivalent   width  of  each
absorption  feature is  measured in  the original  spectrum.  For each
line,  the signal  to noise  ratio  is plotted  as a  function of  the
distance to the central pixel (panel  f). The S/N ratio of the line is
taken at the maximum of the curve.  } }
\label{fig:mdfilt}
\end{figure*}
}

\def\figdamp{
\begin{figure}
\begin{center}
\resizebox{5cm}{!}{\includegraphics{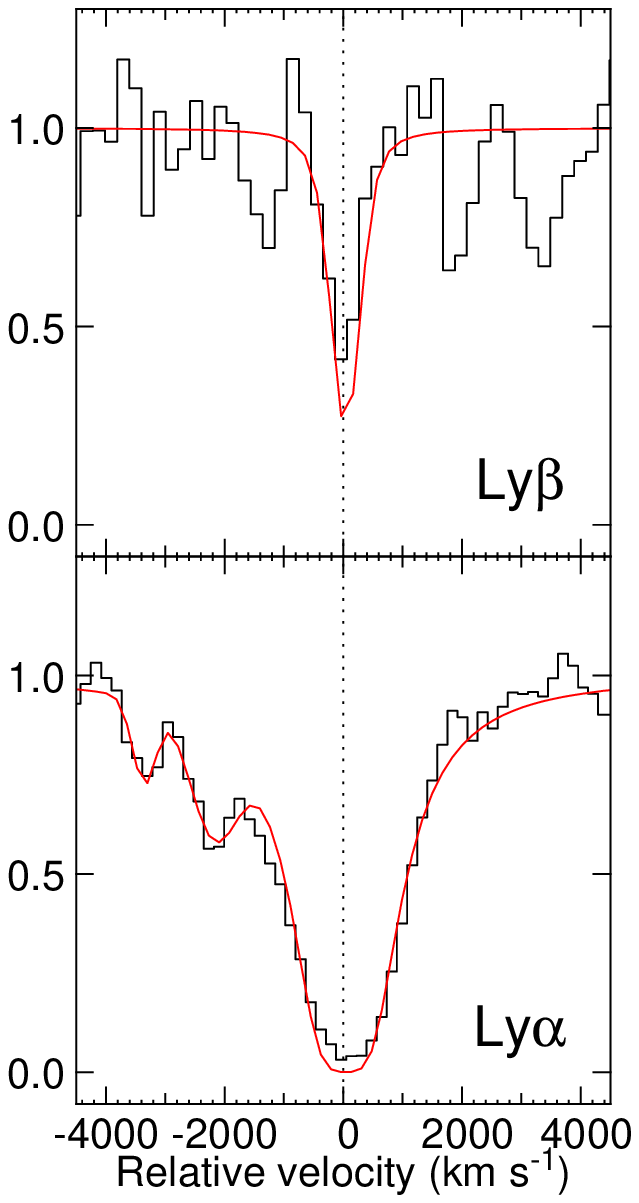}}
\end{center}

\caption{\lya\ and \lyb\  absorptions centered at \zabs{1.2412} toward
\qaa.   The  continuous  curve  is  the  best fit  to  the  two  \Hly\
absorptions obtained with log~$N$(\Hly)~$\ssim$~20.5.  }
\label{fig:fitdamp}
\end{figure}
}

\def\figspec{
\begin{figure*}
\resizebox{\hsize}{!}{\includegraphics{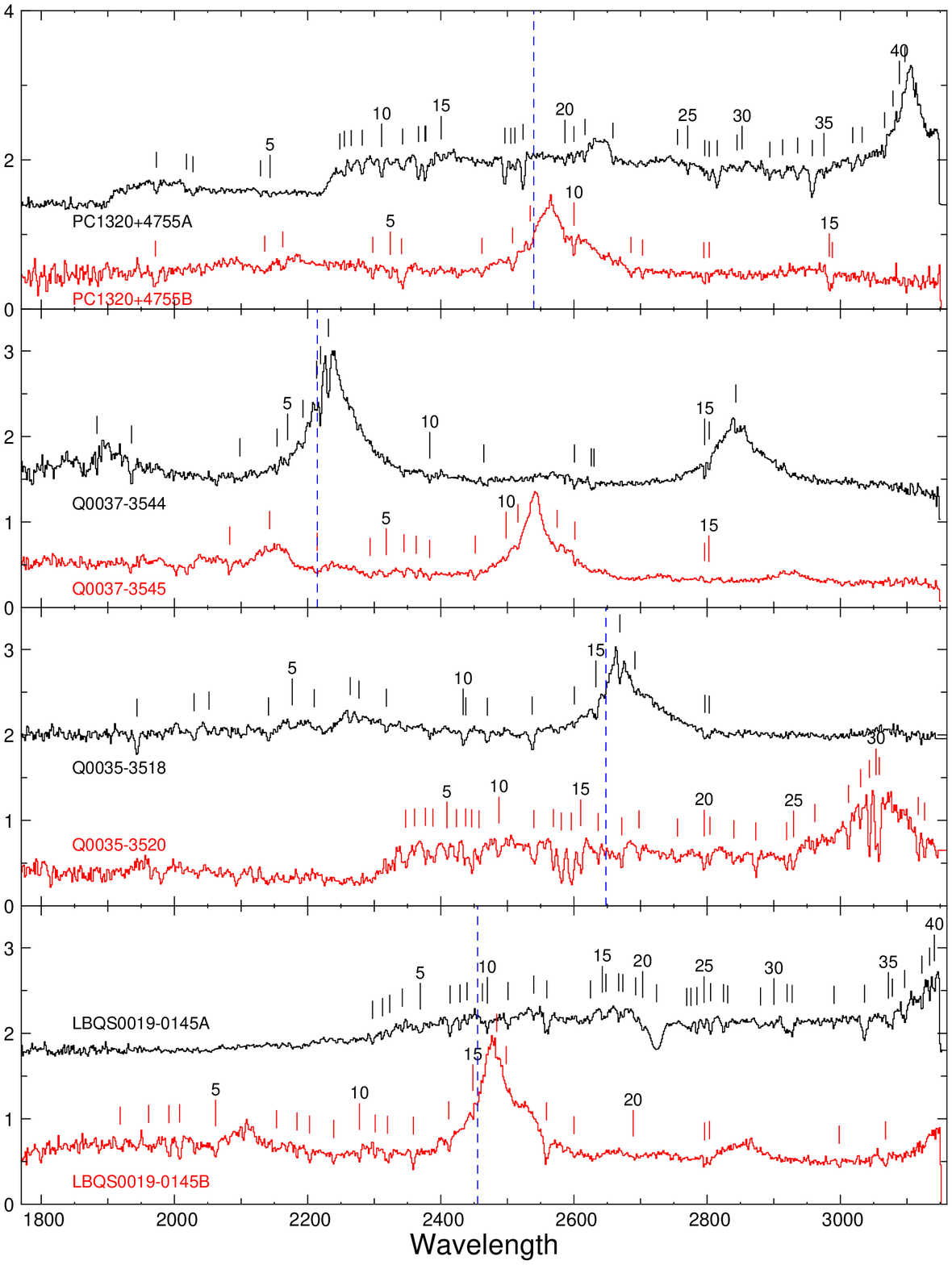}}
\caption{\HST\  STIS spectra  of the  eight observed  \qsos.  The tick
marks  indicate significant  ($>$4$\sigma$)  absorption features.  The
vertical dashed lines  mark the red limit of  the wavelength range for
the  selection  of the  \lya\  lines. This  limit  is  defined by  the
wavelength at which a line lies at more than 3000~km~s$^{-1}$ blueward
of the  \lya\ emission line of  the lowest redshift quasar  in a given
pair.}
\label{fig:spectra}
\end{figure*}
}

\def\figsysA{
\begin{figure}
\resizebox{\hsize}{!}{\includegraphics{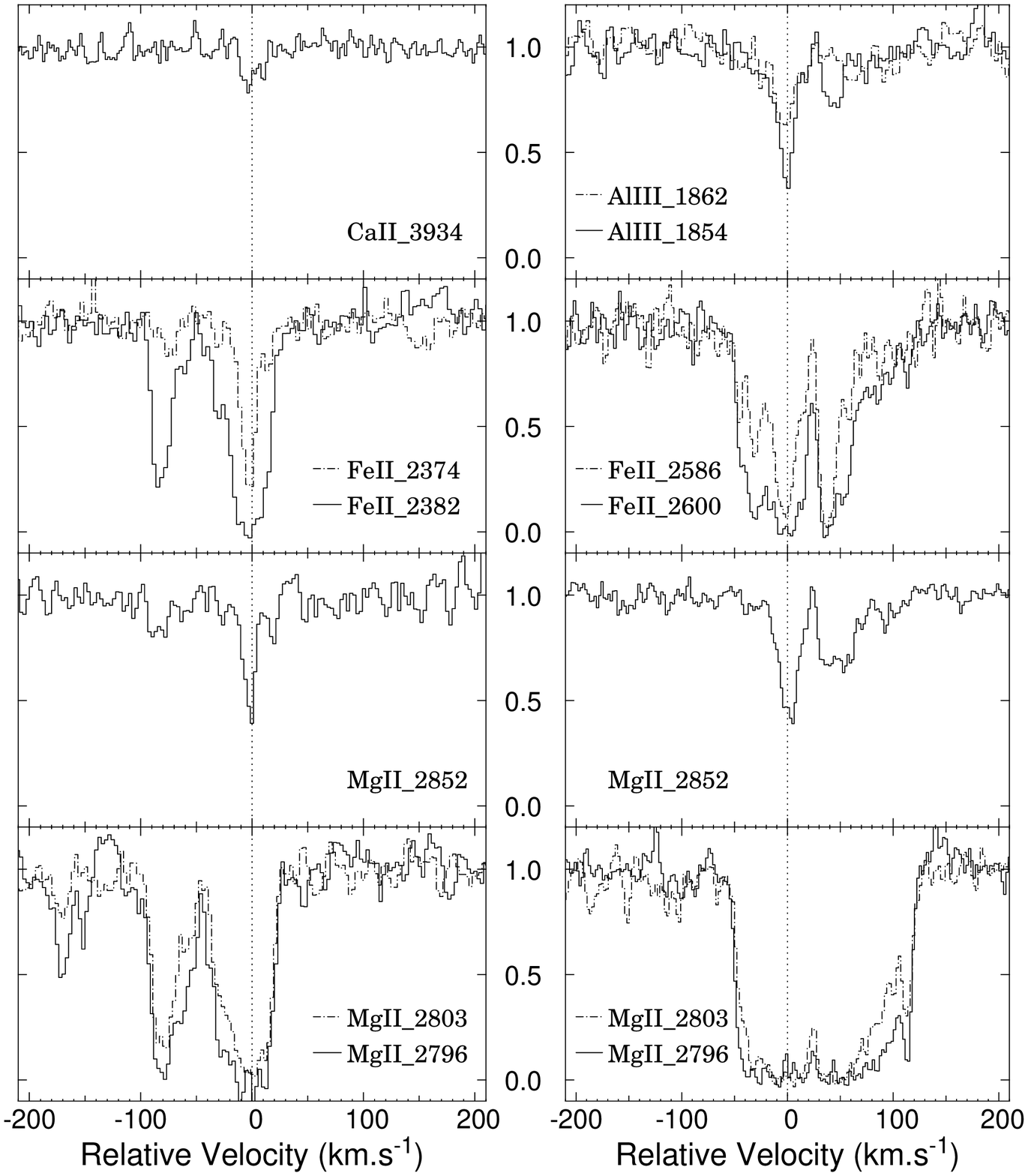}}
\caption{Metal  absorption  lines  at  $z$~=~0.6953  (left  column)  and
$z$~=~1.2412 (right  column, damped Lyman-$\alpha$ system)  in the \qaa\ %
UVES spectrum.}
\label{fig:uvessys}
\end{figure}
}

\def\figcor{
\begin{figure}
\resizebox{\hsize}{!}{\includegraphics{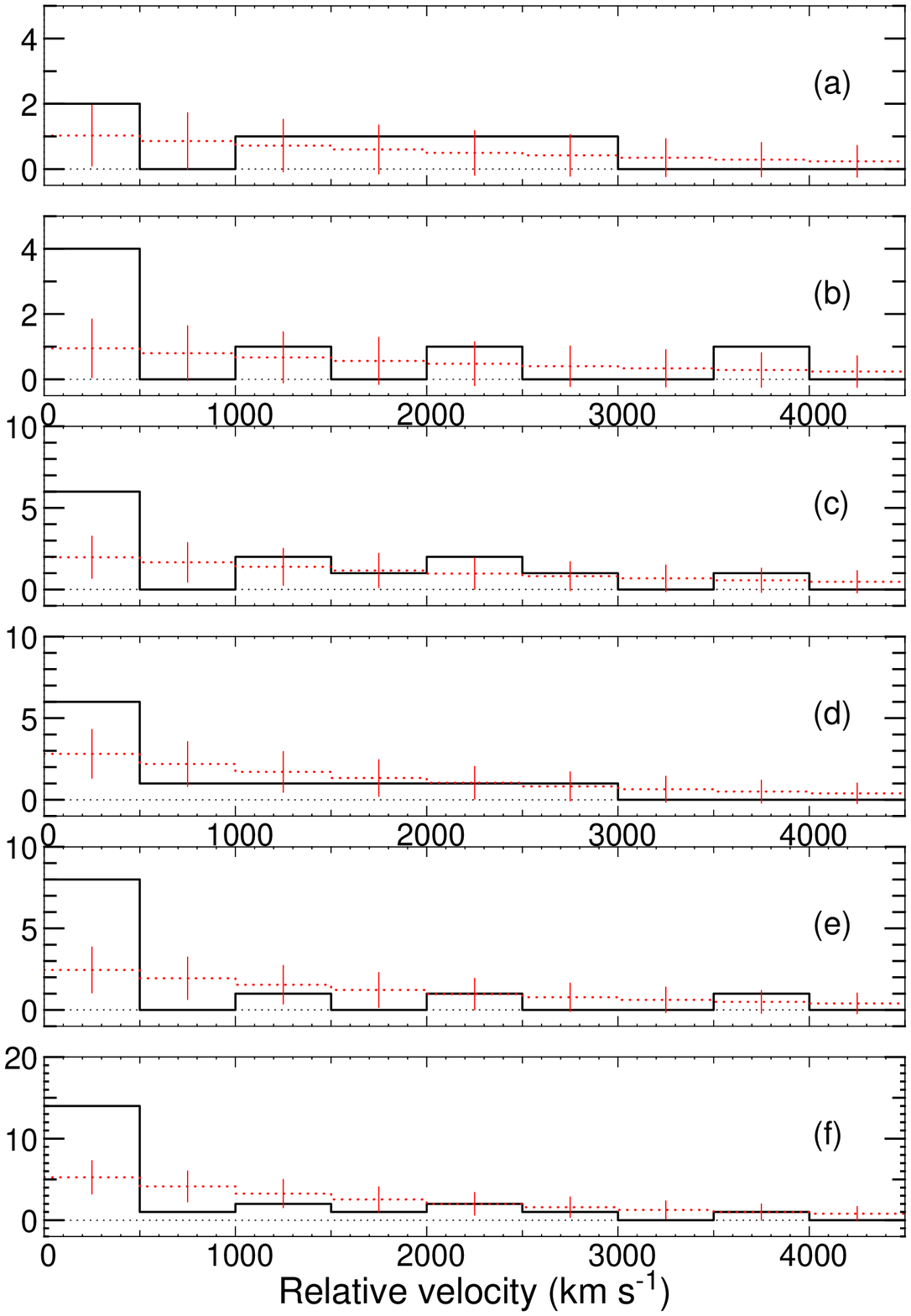}}
\caption{Number of  coincidences versus the  velocity separation \adv\ %
between  the  two  lines   for  rest  equivalent  width  threshold  of
$w_{\rm{r}}>$~0.3~\AA.  {\sl  Panel} (a):  for  pairs with  separation
$\sim\,$2~arcmin,  Q~0037$-$3544  \&  Q~0037$-$3545 and  \qda\&B;  {\sl
Panel}  (b): for  pairs with  separation $\sim\,$3~arcmin,  \qaa\&B and
Q~0035$-$3518  \&  Q~0035$-$3520;  {\sl   Panel}  (c):  for  the  four
pairs. Panels  (d),(e) and (f)  are the same as,  respectively, Panels
(a), (b)  and (c) but after  adding data of  \cite{You01} 2001. Dotted
lines  correspond  to  the  expected  number of  coincidences  from  a
randomly placed population of lines.  }
\label{fig:corr}
\end{figure}
}

\def\figcorI{
\begin{figure}
\begin{center}
\resizebox{7cm}{!}{\includegraphics{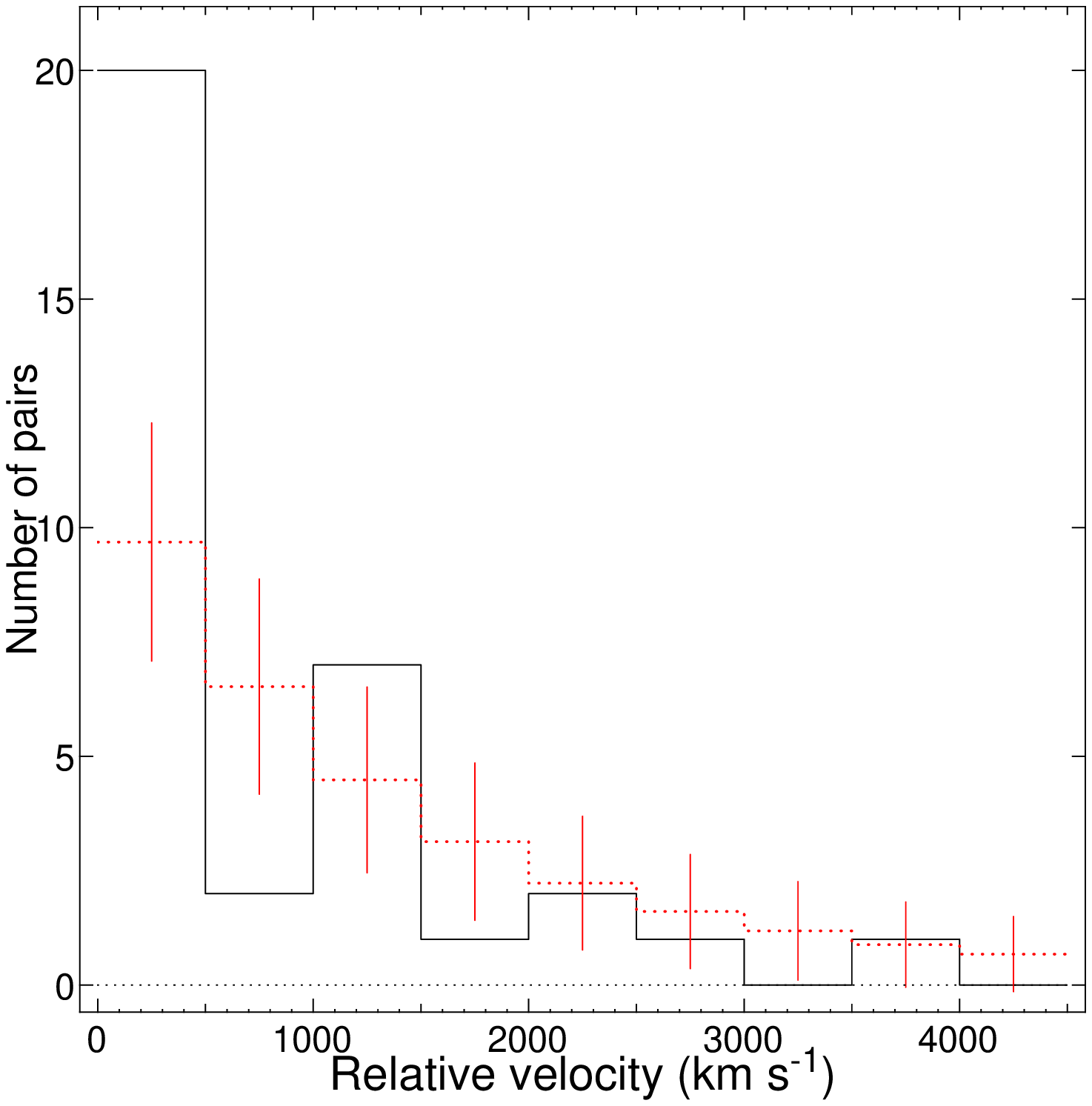}}
\caption{Number of coincidences versus the velocity separation 
\adv\ between the two lines for the complete sample.}

\label{fig:corrI}
\end{center}
\end{figure}
}

\def\figcorIbin{
\begin{figure}
\begin{center}
\resizebox{7cm}{!}{\includegraphics{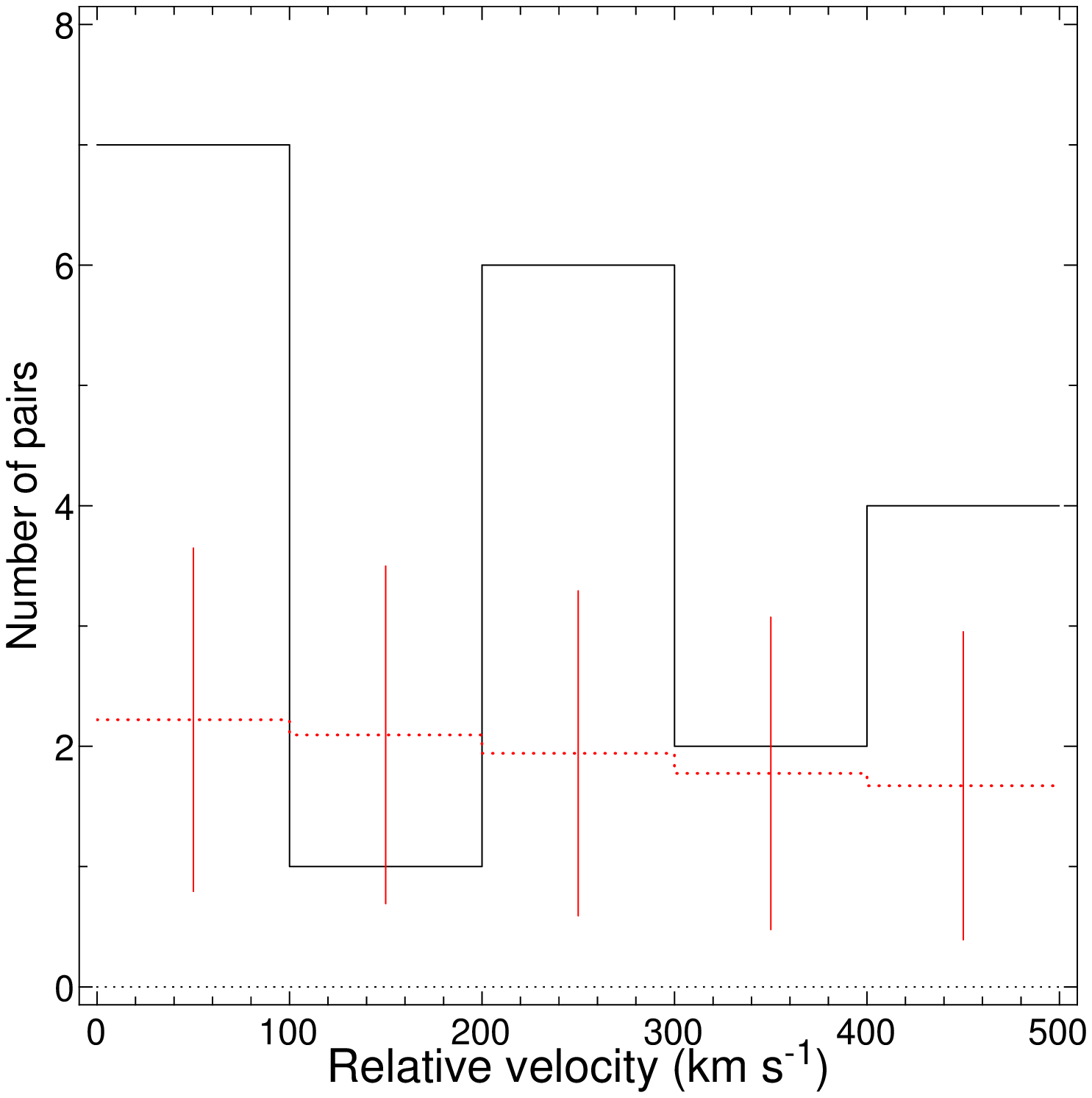}}
\caption{Velocity   distribution   of   the   20   coincidences   with
\adv~$<$~500~km~s$^{-1}$.   The  dotted  line indicates  the  expected
number of coincidences from the Monte-Carlo simulations.}
\label{fig:corr1bin}
\end{center}
\end{figure}
}

\def\figw1vsw2{
\begin{figure}
\begin{center}
\resizebox{7cm}{!}{\includegraphics{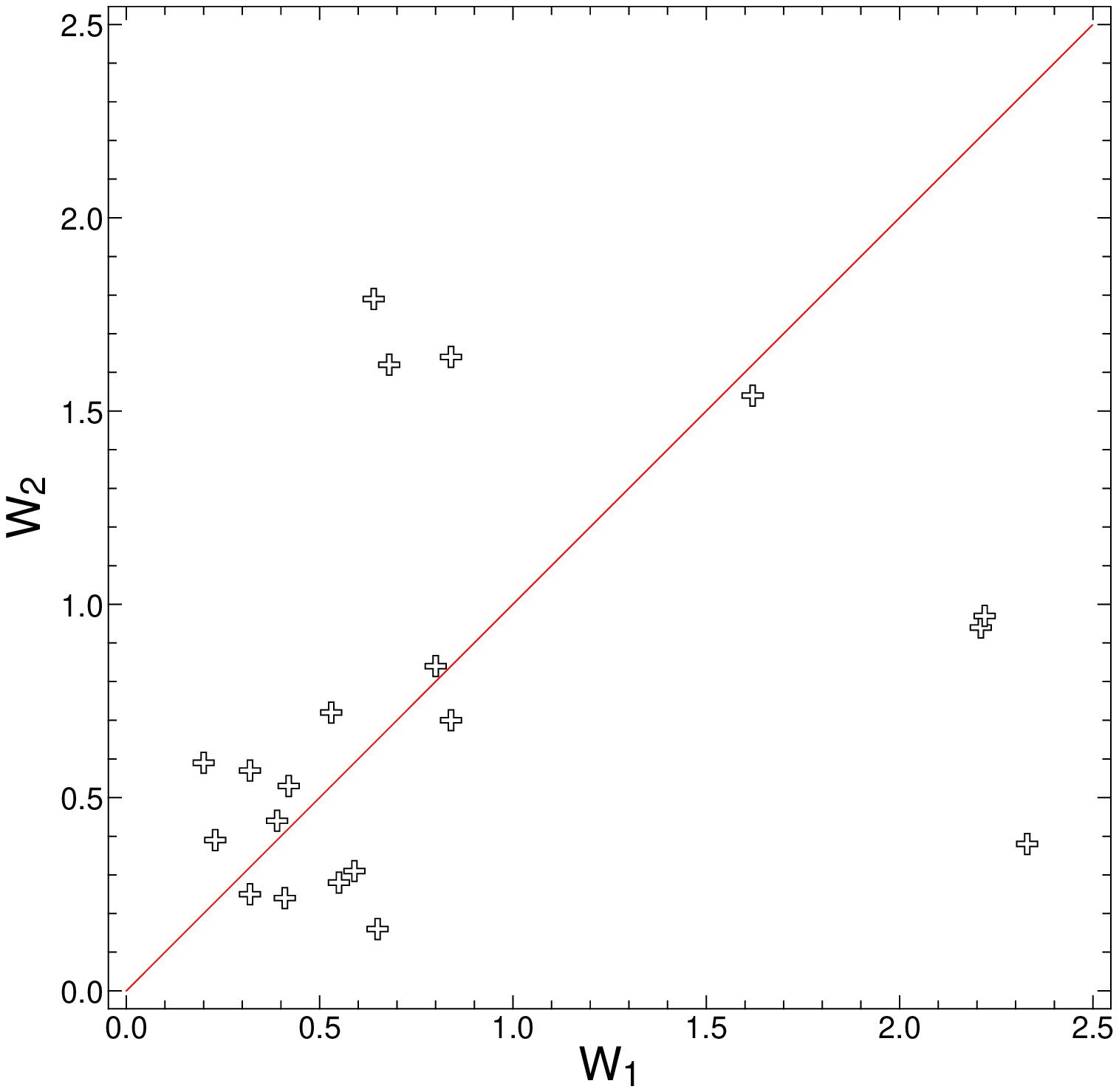}}
\caption{Equivalent width observed along  one line of sight versus the
one observed  along the second line  of sight for  the 20 coincidences
with velocity difference smaller than 500~km s$^{-1}$.}
\label{fig:w1vsw2}
\end{center}
\end{figure}
}

\def\dblambda{\mbox{$\lambda$}}
\newcommand{\transi}[2]{#1\textsc{#2}}
\newcommand{\transil}[3]{\transi{#1}{#2}\dblambda#3}

\def\alIIa{\transil{Al}{ii}{1670}}

\def\alIIIab{\transil{Al}{iii}{1854, 1862}}

\def\cIIIa{\transil{C}{iii}{977}}
\def\cIII{\transi{C}{iii}}
\def\cIIa{\transil{C}{ii}{1334}}

\def\cIVa{\transil{C}{iv}{1548}}
\def\cIVb{\transil{C}{iv}{1550}}
\def\cIVab{\transil{C}{iv}{1548, 1550}}
\def\cIV{\transi{C}{iv}}

\def\caII{\transi{Ca}{ii}}

\def\feIIa{\transil{Fe}{ii}{2600}}

\def\feIIc{\transil{Fe}{ii}{2382}}

\def\feIIe{\transil{Fe}{ii}{2344}}
\def\feII{\transi{Fe}{ii}}

\def\lya{\mbox{Lyman-$\alpha$}}
\def\lyb{\mbox{Lyman-$\beta$}}
\def\lyc{\mbox{Lyman-$\gamma$}}
\def\lyd{\mbox{Lyman-$\delta$}}

\def\lyf{\mbox{Lyman-$\zeta$}}

\def\Hlya{\transil{H}{i}{1215}}
\def\Hlyb{\transil{H}{i}{1025}}

\def\Hlyab{\transil{H}{i}{1025, 1215}}
\def\Hlyabc{\transil{H}{i}{972, 1025, 1215}}
\def\Hlyabcd{\transil{H}{i}{949, 972, 1025, 1215}}
\def\Hly{\transi{H}{i}}

\def\mgIIa{\transil{Mg}{ii}{2803}}
\def\mgIIb{\transil{Mg}{ii}{2796}}
\def\mgIIab{\transil{Mg}{ii}{2796, 2803}}
\def\mgII{\transi{Mg}{ii}}

\def\mgIa{\transil{Mg}{i}{2852}}

\def\nIII{\transi{N}{iii}}

\def\nVb{\transil{N}{v}{1238}}

\def\nV{\transi{N}{v}}

\def\oVIa{\transil{O}{vi}{1031}}
\def\oVIb{\transil{O}{vi}{1037}}
\def\oVI{\transi{O}{vi}}

\def\siIIc{\transil{Si}{ii}{1260}}

\def\siIIed{\transil{Si}{ii}{1190, 1993}}
\def\siII{\transi{Si}{ii}}

\def\siIIIa{\transil{Si}{iii}{1206}}

\def\siIVa{\transil{Si}{iv}{1393}}
\def\siIVb{\transil{Si}{iv}{1402}}
\def\siIVab{\transil{Si}{iv}{1393, 1402}}
\def\siIV{\transi{Si}{iv}}

\def\znII{\transi{Zn}{ii}}

\def\qso{quasar}
\def\qsos{quasars}
\def\qaa{LBQS~0019$-$0145A}
\def\qab{LBQS~0019$-$0145B}
	 
\def\qba{Q~0035$-$3518}
\def\qbb{Q~0035$-$3520}
	 
\def\qca{Q~0037$-$3544}
\def\qcb{Q~0037$-$3545}
	 
\def\qda{PC~1320+4755A}
\def\qdb{PC~1320+4755B}

\def\adv{\mbox{$|\Delta v|$}}
\def\kmps{\mbox{~km~s$^{-1}$}}
\def\radi#1{\mbox{#1~$h_{50}^{-1}$~kpc}}
\def\ssim{\mbox{$\sim$}}
\def\weqr{\mbox{$w_\mathrm{r}$}}
\def\zabs{\mbox{$z_\mathrm{abs}$~=~}}
\def\zem{\mbox{z$_\mathrm{em}$}}
\def\sysabs#1{\zabs #1}
\def\wvl{wavelength}
\def\wvls{wavelengths}
\def\stis{Space Telescope Imaging Spectrograph}
\def\parn{\par\noindent}
\def\aeta{A\&A}

\def\aetas{A\&AS}
\def\apj{ApJ}

\def\mn{MNRAS}

\def\los{line of sight}
\def\loss{lines of sight}
\def\ew{equivalent width}

\def\ablns{absorption lines}

\def\HST{{\it HST}}
\def\hst{{\it Hubble Space Telescope}}

\def\subsec#1{\vspace{-0.3cm}\subsection{#1}\vspace{0.4cm}}
\def\subsubsec#1{\vspace{-0.4cm}\subsubsection{#1}\vspace{-0.25cm}}

\begin{document}

\title{{\HST}  STIS  observations of  four  QSO pairs\thanks{Based  on
observations with  the NASA/ESA \hst, obtained at  the Space Telescope
Science   Institute,  which   is  operated   by  the   Association  of
Universities  for Research  in  Astronomy, Inc.,  under NASA  contract
NAS5-26555.   Based  on  observations  carried  out  at  the  European
Southern Observatory (ESO, programme  No.  66.A-0624) with UVES on the
8.2~m VLT-Kuyen telescope operated at Paranal Observatory; Chile.  }}

%ESO 66.A-0624

\titlerunning{ {\it HST} STIS observations of four QSO pairs}

\author{ Bastien Aracil\inst{1}
   \and  Patrick Petitjean\inst{1,2}
   \and  Alain Smette\inst{3,4}
   \and  Jean Surdej\inst{4,5}
   \and  Jan Peter M\"ucket\inst{6}
   \and  Stefano Cristiani\inst{7,8}
}

\offprints{Bastien Aracil (aracil@iap.fr)}

\institute{$^1$ Institut d'Astrophysique de Paris - CNRS,
                98bis Boulevard Arago, F-75014 Paris, France\\
           $^2$ LERMA, Observatoire de Paris, 61 Avenue de l'Observatoire,
                F-75014 Paris, France\\
           $^3$ Chercheur qualifi\'e et Collaborateur Scientifique,
                Fonds National de la Recherche
                Scientifique, Belgium\\
           $^4$ Institut d'Astrophysique et de G\'eophysique,
                Universit\'e de Li\`ege, Avenue de Cointe 5,
                B-4000 Li\`ege\\
           $^5$ Directeur scientifique, Fonds National de la Recherche
                Scientifique, Belgium\\
           $^6$ Astrophysikalisches Institut Potsdam,
                An der Sternwarte, Potsdam, Germany\\
           $^7$ Space Telescope European Coordinating Facility,
                Karl-Schwarzschild Str. 2, D-85748 Garching, Germany\\
	   $^8$ Osservatorio Astronomico di Trieste, Via G.B. Tiepolo, 11
 	        I-34131 Trieste, Italy}

\date{Received ?? / Accepted ?? }

\abstract{  We present \HST\  STIS observations  of four  quasar pairs
with redshifts 0.84~$<$~$z_{\rm em}$~$<$~1.56 and angular separation 2--3~arcmin
corresponding  to  $\sim\,$1$-$1.5$h^{-1}_{50}$~Mpc  transverse  proper
distance  at $z$~$\sim$~0.9.   We  study  the  distribution  of
velocity  differences between  nearest  neighbor \Hly\  Lyman-$\alpha$
absorption lines detected in the  spectra of adjacent QSOs in order to
search  for the  possible  correlation  caused by  the  extent or  the
clustering properties of the structures traced by the absorption lines
over  such a  scale. The  significance  of the  correlation signal  is
determined by comparison with  Monte-Carlo simulations of spectra with
randomly  distributed absorption lines.   We find  an excess  of lines
with a  velocity separation smaller  than $\Delta V$~=~500~km~s$^{-1}$
significant  at  the  99.97\%  level.  This  clearly  shows  that  the
Lyman-$\alpha$   forest   is   correlated   on  scales   larger   than
1$\,h^{-1}_{50}$~Mpc  at   $z$~$\sim$~1.   However,  out   of  the  20
 detected coincidences   within    this   velocity   bin,    12   have   $\Delta
V$~$>$~200~km~s$^{-1}$. This probably reflects the fact that the scale
probed  by  our  observations is  not  related  to  the real  size  of
individual   absorbers  but  rather   to  large   scale  correlation.
Statistics  are too  small to  conclude about  any  difference between
pairs  separated by  either 2  or 3~arcmin.   A  damped Lyman-$\alpha$
system is detected  at $z_{\rm abs}$~=~1.2412 toward LBQS~0019$-$0145A
with log~$N$(\Hly)~$\sim$~20.5. From the absence of \znII\ absorption,
we derive a metallicity relative to solar [Zn/H]~$<$~$-$1.75.
}

\maketitle

\section{Introduction}
Recent  $N$-body  numerical  simulations  reproduce  successfully  the
global characteristics of the neutral hydrogen absorptions observed in
quasar  spectra,  the  so-called Lyman-$\alpha$  forest  (\cite{Cen94}
1994;  \cite{PPj95}   1995,  \cite{Her96}  1996;   \cite{Zha95}  1995;
\cite{Muc96}   1996;  \cite{Mir96}   1996;  \cite{Bon98}   1998).  The
absorptions   arise   from  density   inhomogeneities   in  a   smooth
all-pervading   intergalactic  medium.   Simulations  show   that  the
intergalatic gas traces the potential wells of the dark matter well at
high   redshift.   It   is   therefore  possible   to  constrain   the
characteristics of  the dark-matter density field  from observation of
the  Lyman-$\alpha$ forest  along a  single  line of  sight (Croft  et
al. 2000). The addition  of transverse information from observation of
QSO  pairs or  more generally  groups  of quasars  at small  projected
separation in  the sky will  probably revolutionize this field  in the
next  few years  (\cite{PPj97} 1997).  Indeed, inversion  methods that
have been recently implemented show that it is possible to recover the
3D topology of the dark-matter field using a dense network of lines of
sight (Nusser \& Haehnelt 1998, Pichon et al. 2001, Rollinde et al. 2001).
\par\noindent
After the  early discovery of  common absorptions in pairs  of quasars
(Shaver et al.  1982, Shaver \& Robertson 1983,  Weyman \& Foltz 1983,
Foltz et  al. 1984),  it was shown  that the gaseous  complexes giving
rise to  the absorptions should have large  dimensions. In particular,
studies  of  gravitationally  lensed  quasars  (Smette  et  al.  1992,
\cite{Sme95}  1995)  yielded  a  lower  limit of  \radi{100}\  on  the
diameter of \lya\ absorbers.  Similar results were obtained from pairs
of \qsos\ with small separation (\cite{Bec94} 1994; \cite{Din94} 1994;
Petitjean et al. 1998,  \cite{VDo98} 1998; \cite{Mon98} 1998).  Larger
separations   have   been   investigated   by   \cite{Cro98}   (1998),
\cite{Din98} (1998),  \cite{Mon99} (1999) and Williger  et al. (2000).
All studies conclude that  absorptions are correlated on scales larger
than \radi{500}.
\par\noindent
Unlike the case of QSO  pairs with small angular separations where the
correlation can be explained by the fact that the \loss\ intercept the
same absorber, the correlation for larger separations is certainly due
to  the  clustering  properties  of distinct  clouds.  When  observing
triplets  of  quasars  separated  by   1  to  2  arcmin  on  the  sky,
corresponding to  $\sim\,$0.5 to 1~Mpc proper distance scales, both  Crotts \& Fang  (1998) and
Young et al. (2001) find statistically significant triple coincidences
that  they interpret  as the  presence of  sheetlike  structures along
which inhomogeneous absorbers cluster.
\par\noindent
The number of such experiments is small  however and it is important to
increase the  statistics. Here we  present \HST\ observations  of four
pairs  of \qsos.  The $\sim\,$2  and $\sim\,$3  arcmin  angular separations
between the  two quasars  of each pair  probes scales between  1.0 and
1.5~$h^{-1}_{50}$~Mpc proper distance at  $z$~$\sim$~1.  This is where the
transition  between  individual halos  and  filamentary or  sheet-like
large  scale structures is  expected (\cite{Muc96}  1996, \cite{Cha97}
1997).
\par\noindent
We  describe the  observations in  Sect.~2 and  comment  on individual
metal  line systems in  Sect.~3. Correlations  between metal  line and
Lyman-$\alpha$  systems  are  respectively  discussed in  Sect.~4  and
5. Conclusions are drawn in Sect.~6.
\section{Observations}\label{sec:obs}
\figspec
\figfilter
\begin{table}[h]
\caption{List of observed QSOs}
\begin{small}
\begin{minipage}{\hsize}
\renewcommand{\footnoterule}{}
\begin{tabular}{l c c c c}
\hline
Object name    & z$_{\rm em}$
               & $\Delta\theta$\footnote{\small Angular separation on the sky in arcmin}
               & $\Delta z$\footnote{\small Redshift range over which coincidences were searched
for}
               & $<\rm{S}>$\footnote{\parbox[t]{8cm}{\small Mean proper distance in kpc between 
lines of sight in the redshift range ($q_0=0.5, \Lambda=0,
H_0=50$~km~s$^{-1}$~Mpc$^{-1}$)}}\\
\hline
LBQS0019-0145A & 1.59     &                &                 &                     \\
LBQS0019-0145B & 1.04     & 3.3            & $0.70\,-\,1.01$ & 1640                \\
Q0035-3518     & 1.20     &                &                 &                     \\
Q0035-3520     & 1.52     & 3.4            & $0.69\,-\,1.17$ & 1710                \\
Q0037-3545     & 1.10     &                &                 &                     \\
Q0037-3544     & 0.84     & 1.7            & $0.59\,-\,0.82$ &  810                \\
PC1320+4755A   & 1.56     &                &                 &                     \\
PC1320+4755B   & 1.11     & 1.9            & $0.62\,-\,1.08$ &  940                \\
\hline
\end{tabular}
\vspace{-0.25cm}
\end{minipage}
\end{small}
\vspace{-0.5cm}
\end{table}
\par\noindent
Observations were  carried out  on the \hst\  using the  \stis\ (STIS)
with  the   G230L  grating   and  the  Near-UV-MAMA   detector.   This
configuration yields a mean spectral resolution of R = 700 (FWHM = 3.4
\AA\ at $\lambda$  = 2374~\AA) and a \wvl\ coverage  from 1570~\AA\ to
3180~\AA.  The  observations were reduced at the  Goddard Space Flight
Center with  the STIS Investigation  Definition Team (IDT)  version of
CALSTIS (\cite{Lin98}  1998). Standard reduction  and calibration were
used.  Special  care was taken to determine  accurately the background
due to the sky  and the dark  current.  The  zero point of  the \wvl\
scale for  individual exposures  was determined requiring  the Galatic
interstellar absorptions to occur  at rest.  The correction can always
be performed because  the Galactic \mgII\ doublet is  well detected in
every single  spectrum.  When the  \feII\ lines were also  detected we
checked that  the dispersion in the zero point is smaller than  the spectral resolution.
The resulting  spectra are shown  in Fig.~1.  The \qso\  continuum was
fitted  with Gaussian  profiles for  emission lines  and  simple cubic
splines in regions  between emission lines. The  best fit was found
by varying  the position of the  control points of  the cubic splines.
The resulting continuum was slightly manually adjusted in regions that
were poorly fitted  as for example near broad  emission lines and Lyman
limits.  The detection of  absorption lines in the normalized spectrum
was  performed by  filtering  each spectrum  to  improve the  contrast
between  lines and  noise.  This  has been  performed  by successively
applying a  wavelet filter  and an ``upgraded''  median filter  to the
spectrum.
\par\noindent
We  used for  the  wavelet filter  B3-spline  scaling functions.  This
filter selects the wavelength scales corresponding to the width of the
\ablns~  (see  panel b  of  Fig.~\ref{fig:mdfilt}).   Pixels from  the
wavelet filtered spectra are  sorted in increasing order keeping trace
of the  pixel permutations.  The distribution of  the pixel  values is
shown in  panel (c)  of Fig.~\ref{fig:mdfilt}.  A  lower limit  of the
level  of noise  in  each pixel  can  be estimated  (dashed line)  and
substracted to the  real pixel value (panel d  in Fig.~2).  Pixels are
then reordered  and the  resulting spectrum is  shown in panel  (e) of
Fig.~\ref{fig:mdfilt}.  This  filtered  spectrum  is  used  to  define
regions where possible absorption  lines are present using a threshold
defined so that no  line is lost in the next step.  We then compute in
the original  spectrum the equivalent  width and the  associated noise
over each  of these regions and  select only those  with an equivalent
width to noise ratio larger than 2. The corresponding regions are then
fitted with a Voigt profile  fitting program to derive the position of
absorption features.  This software  makes a $\chi^2$  minimisation in
each region adding lines until the reduced $\chi^2$ reaches a value lower than
or equal to 1.
\par\noindent
For  each absorption  feature, we  calculate the  equivalent  width in
windows  centered  on  the  minimum  of  the line  and  of  widths  an
increasing number of pixels.   The signal-to-noise ratio computed from
the noise  spectrum is plotted  as a function  of the distance  to the
central pixel (panel f of Fig.~2).  The S/N
ratio  of the  line was  taken at  the maximum  of the  curve  and the
equivalent width  was computed by integrating the  fitted profile. The
lines  with   S/N  ratio  greater   than  4  are  listed   with  their
identification in Tables 2 to  5. In these tables, uncertain positions
or identifications are indicated by a colon.
\section{Comments on individual metal line systems}
\begin{table}
\caption{Line list for the pair LBQS0019-0145AB}\label{tab:qA}
\resizebox{\hsize}{!}{\includegraphics{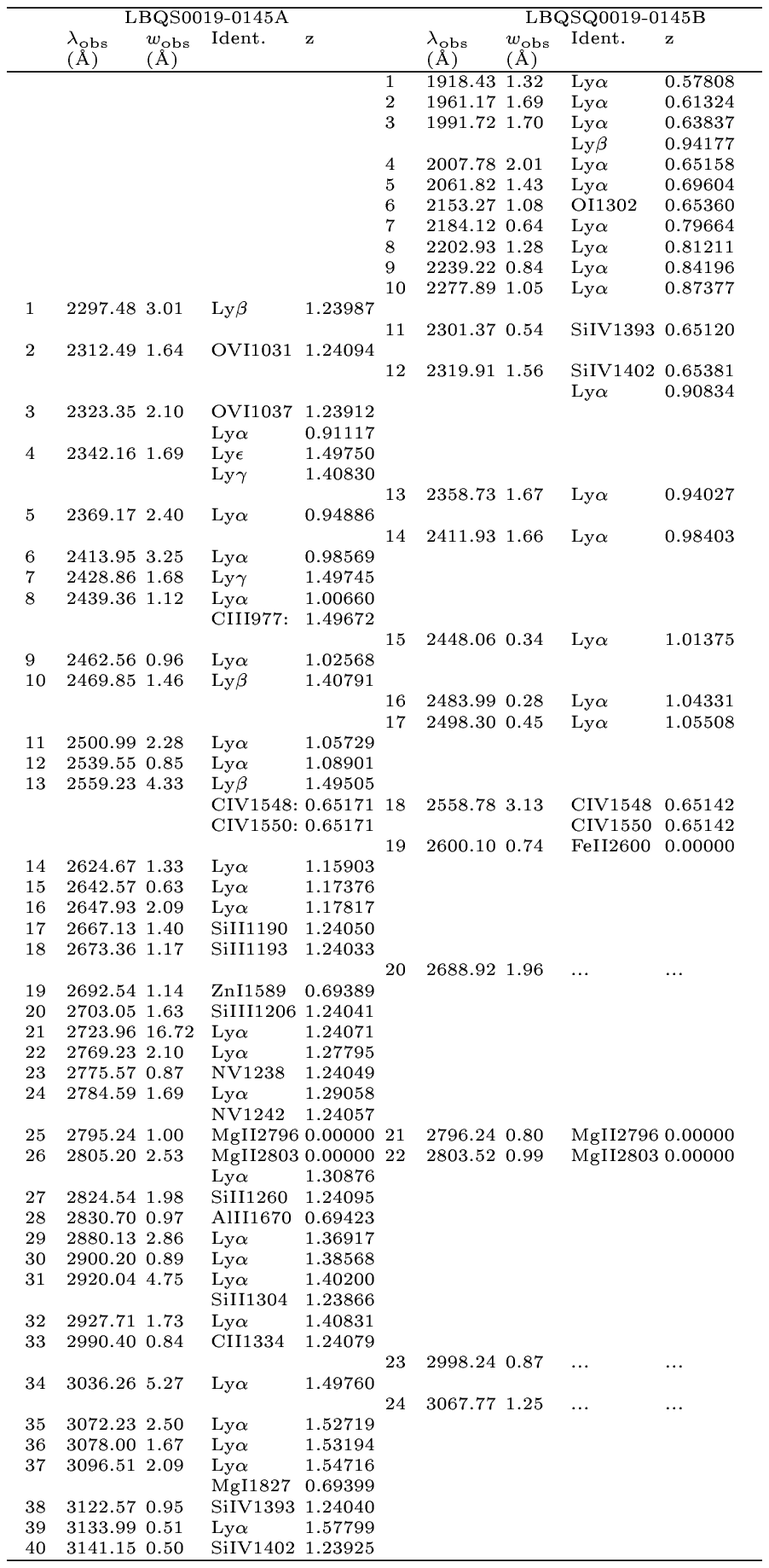}}
\end{table}
\begin{table}
\caption{Line list for the pair LBQS0035-3518 \& Q0035-3520}\label{tab:qB}
\resizebox{\hsize}{!}{\includegraphics{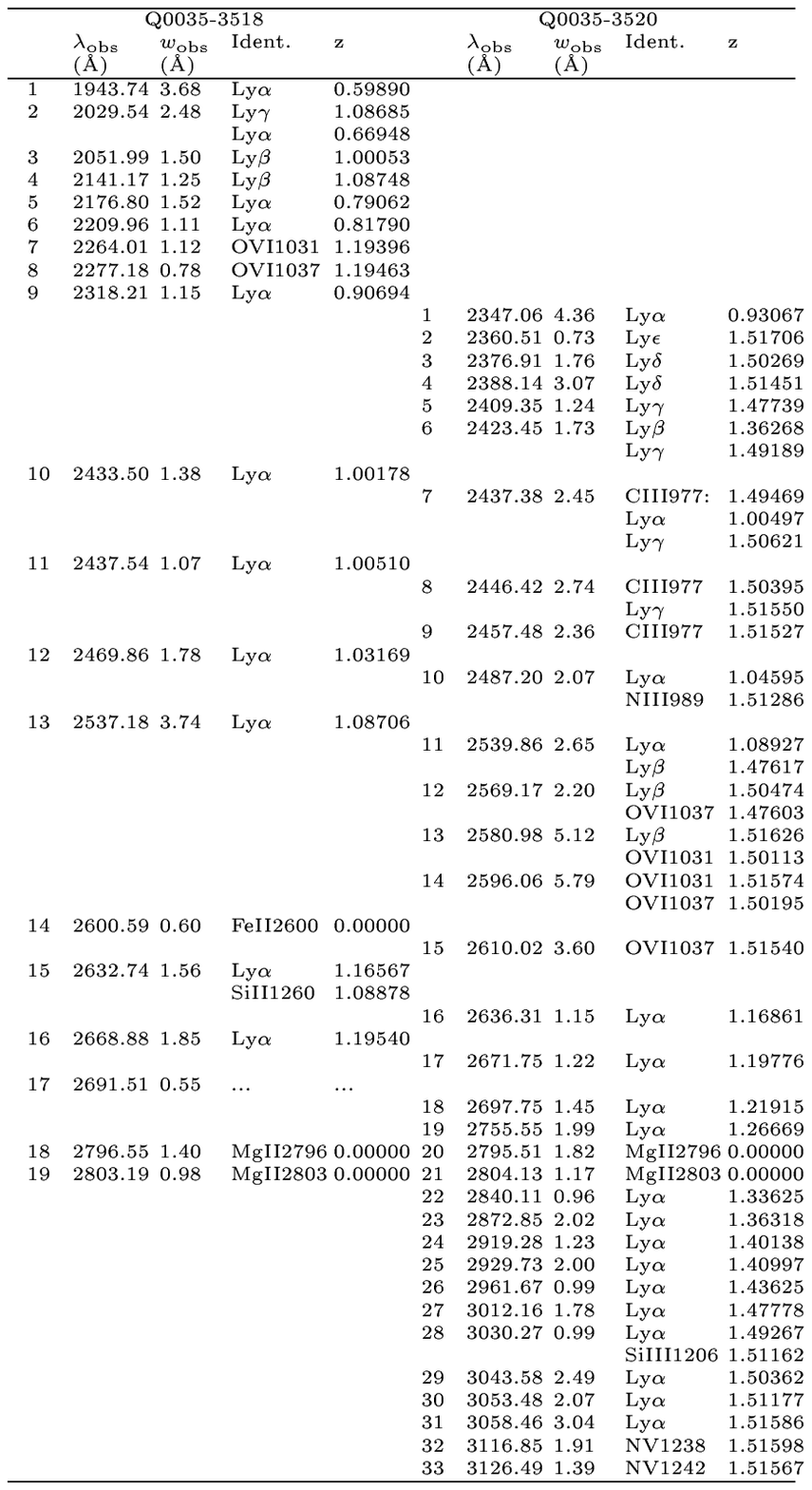}}
\end{table}
\begin{table}
\caption{Line list for the pair Q0037-3544 \& Q0037-3545}\label{tab:qC}
\resizebox{\hsize}{!}{\includegraphics{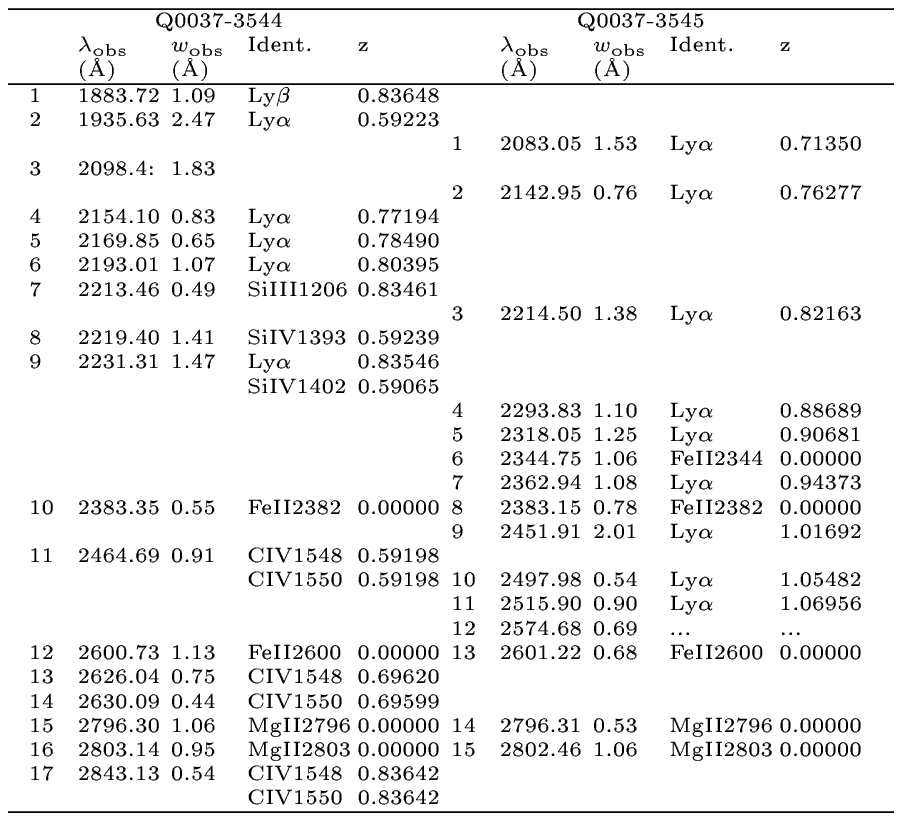}}
\end{table}
\begin{table}
\caption{Line list for the pair PC1320+4755AB}\label{tab:qD}
\resizebox{\hsize}{!}{\includegraphics{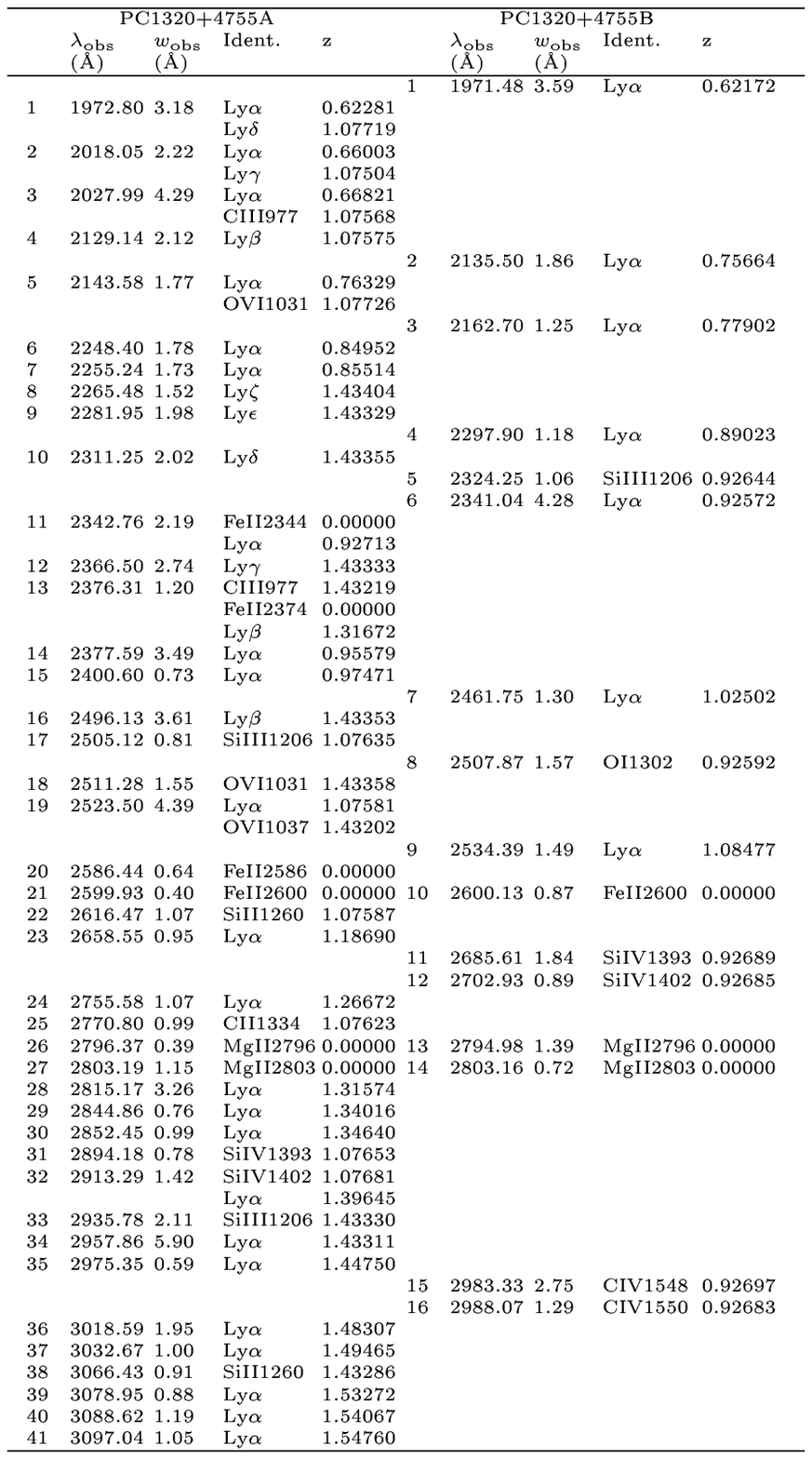}}
\end{table}
%%
%%
%% System de LBQS0019A
%%
%%
In  this Section  we comment  on  the intervening  metal line  systems
identified in the  spectra. To refer to an  absorption feature, we use
the numbering given in Tables~2 to 5.
\subsection{\qaa\ ~~\zem~=~1.59}
The  \HST\ data  on this  quasar have  been complemented  with  a high
resolution (R~$\sim$~40000) UVES spectrum covering
the wavelength ranges 3900$-$5200~\AA\ and 5450$-$9300~\AA. The exposure
time was two hours.
\subsubsec{$z_{\rm abs}$~=~0.6514}
The  strong  absorption feature  \#13  at  $\lambda$2559.23 cannot  be
explained by Lyman-$\beta$ at $z_{\rm abs}$~=~1.4976 alone because the
corresponding \ew\ is too large.  As there is a strong \cIV\ system at
$z_{\rm abs}$~=~0.65142  in the spectrum of \qab,  we tentatively note
that  the additional  absorption  could  be due  to  \cIV\ at  $z_{\rm
abs}$~=~0.6514.    The  corresponding  Lyman-$\alpha$   absorption  is
unfortunately  redshifted  below  the   Lyman  limit  of  the  $z_{\rm
abs}$~=~1.4976 system.  \mgIIa\ at this redshift is not  detected in 
the  UVES spectrum down to $w_{\rm obs}=35$~m\AA.
\subsubsec{\sysabs{0.6953}}
\figsysA%%
Mg{\sc ii}$\lambda$2796 (\weqr\ssim0.8~\AA), \feIIc\ (\weqr\ssim0.5~\AA),
\mgIa\ and  \caII\ absorptions  are detected at  this redshift  in the
UVES spectrum in addition to  \alIIa\ (line \#28) which is detected in
the \HST\  spectrum. Unfortunately, the  \transil{H}{i}{1215}\ line is
redshifted  at $\lambda$  = 2060~\AA,  below  the Lyman  limit of  the
$z_{\rm abs}$~=~1.4976 system.
\parn
The profile of the \mgII\  and \feII\ absorptions consists of two main
absorption features  approximately $100\,\mathrm{km~s^{-1}}$ apart and
separated  by a  sharp  drop in  optical  depth near  the center  (see
Fig.~\ref{fig:uvessys}).   As described  by  \cite{Bon01} (2001),  and
despite  the  moderate  \mgIIb\  equivalent width,  this  profile  may
indicate that the line of sight intercepts a superwind.
\subsubsec{\sysabs{1.2412}}
\figdamp
Strong  \Hlya\ (line  \#21), \Hlyb,  \siII, \siIIIa,  \siIV,  \nV\ and
\cIIa\  absorptions associated with  this system  are detected  in the
\HST\  spectrum.   Additional  \mgII,  \mgIa,   \feII\  and  \alIIIab\
absorptions     are    seen    in     the    UVES     spectrum    (see
Fig.~\ref{fig:uvessys}).   Despite  the low  resolution  of the  \HST\
spectrum,  damped  wings   are  clearly  seen  in  the   case  of  the
Lyman-$\alpha$  line. The  simultaneous  fit of  the  \lya, and  \lyb\
absorptions gives  log~$N$(\Hly)\ssim20.5 (see Fig.~4).   The observed
\feII\ and  \mgII\ lines  are heavily saturated  so that only  a lower
limit  on  the column  densities  can  be  derived, log~$N$(\feII\  \&
\mgII)$>$15.       \znII\       is       not       detected       with
log~$N$(\znII)~$<\nolinebreak11.4$ implying  that metallicity relative
to solar, [Zn/H]~$<$~$-$1.75,  is one of the smallest  observed at low
redshift  (see e.g.  \cite{led01}  2001). Note  that the  large column
density found for  \mgII\ implies that the neutral  part of the system
cannot account for all the \mgII\ we see.
\parn
As for  the system at  \zabs{0.6953}, the profiles of  the absorptions
consist of two main components  separated by about 50\kmps. This again
may indicate that the \los\ intercepts a superwind.
\subsubsec{\sysabs{1.3088}} 
The system  is unambiguously identified by  sharp \mgIIab\ absorptions
(\weqr\ssim0.07~\AA)  detected  in   the  UVES  spectrum.   The  \lya\
absorption is blended with \mgIIb\ from the interstellar medium.
\subsubsec{\sysabs{1.4976}}   
This  strong system  is detected  by \transil{H}{i}{1215},  1025, 972,
949, 937  and \cIIIa\ absorptions.  It is at  the origin of  the Lyman
limit at  $\lambda$~$\sim$~2270~\AA~ and therefore log~$N$(\Hly)~$>$~18.
A  possible \mgIIb\ absorption line is  present  in the  UVES  spectrum with  $w_{\rm
obs}$~$\sim$~25~m\AA.
%%
%% 
%% System de LBQS0019B
%%
%%
\subsec{\qab\ ~~\zem~=~1.04}
Only  one  metal line  system  along this  line  of  sight  is detected at
  $z_{\rm abs}$~=~0.6513 by \Hlya, \cIVab\ and \siIVab\ absorptions.
%%
%%
%% System of Q0035-3518
%%
%%
\subsec{\qba\ ~~\zem~=~1.20}
%
%\subsubsec{\sysabs{0.0000}}
%The \mgIIab\ doublet and \feIIa absorption are seen fromm the interstellar medium.
%
\subsubsec{\sysabs{1.088}}
Strong \Hlyabc\  and \siIIc\ absorptions are detected  in this system.
The latter line is probably  blended with an other Lyman-$\alpha$ line
however.  Moreover, an absorption  feature detected at the 2.5$\sigma$
level   is  observed   at   the  expected   position   of  \cIIa\   at
$\lambda\ssim2785$~\AA.

\subsubsec{\sysabs{1.1954}}
This  system   is  at  slightly   larger  redshift  than   the  quasar
($\sim\,$+1200~km~s$^{-1}$).   It    shows   strong   associated   \oVI\
absorption. \nVb\ is detected at the 2.5$\sigma$ level.

%%
%%
%% System of Q0035-3520
%%
%%
%\vspace{-0.3cm}
\subsection{\qbb\ ~~\zem~=~1.52}
The  four metal line  systems detected  in this  \qso\ are  all within
3500~km~s$^{-1}$ from  the emission redshift  of the quasar.   As they
are  approximately  at  the  same  redshift,  their  \lyb\  and  \oVI\
absorptions are  blended. The resulting blend corresponds  to the four
strong     features     seen     at     $\lambda\ssim$2600~\AA~     in
Fig.~\ref{fig:spectra}.

\subsubsec{\sysabs{1.4927}}
The \lyb\ line of  this system is observed at $\lambda\ssim$2550~\AA,
but is  below the 4$\sigma$  detection limit. As no  other metal  line
 is  detected,  the \cIIIa\
identification is tentative and  the feature at $\lambda$2437.38 could
be  a   blend  of   Lyman-$\gamma$  at  $z_{\rm   abs}$~=~1.50621  and
Lyman-$\alpha$ at $z_{\rm abs}$~=~1.00492.
\subsubsec{\sysabs{1.50362}}
This  system consists of  \Hlyabcd, \oVI\  and \cIIIa\  absorptions. A
feature  is also  observed at  the  expected \wvl\  of the  associated
\siIIIa\ absorption ($\lambda\ssim$3020~\AA).

\subsubsec{\sysabs{1.5118}}
\Hlyab, \oVI, \cIIIa\ and \siIIIa\ are observed at this redshift. 

\subsubsec{\sysabs{1.5158}}
This system has  all the characteristics of an  associated system with
very  strong \nV\  and \oVI\  lines. Therefore  the \nIII\  and \cIII\
identifications are tentative.
%%
%%
%% System of Q0037-3544
%%
%%
\subsec{\qca\ ~~\zem~=~0.84}

\subsubsec{\sysabs{0.5922}}
Strong  \Hlya\  and  \cIV\  absorptions  are  detected.  \siIVab\  are
redshifted  at   $\lambda$2219.4  and  2231.31~\AA~   (lines  \#8  and
\#9). Given the strengths of the corresponding absorption features, it
is most probable that the two \siIV\ lines are blended with additional
Lyman-$\alpha$  lines.   An  absorption  feature is  detected  at  the
expected  position of \siIIIa\  ($\lambda\ssim$1920~\AA) but  is below
the 4$\sigma$ threshold.

\subsubsec{\sysabs{0.6961}}
This system  consists of  a \cIV\ doublet  detected outside  the \lya\
forest.  The  \lya\ absorption is  observed at \wvl\ 2062~\AA,  but is
below the 4$\sigma$ threshold.

\subsubsec{\sysabs{0.8364}}
\Hlya, \siIIIa\  and \cIV\ absorptions are observed  at this redshift.
Moreover, a  feature is observed  at the expected position  of \siIIc\
($\lambda\ssim$2316~\AA).
%%
%%
%% System of Q0037-3545
%%
%%
\subsection{\qcb\ ~~\zem~=~1.10}
We do not identify any metal line system along this line of sight.

%%
%%
%% System of PC1320+4755A
%%
%%
\subsec{\qda\ ~~\zem~=~1.56}

\subsubsec{\sysabs{1.0762}}
Strong  \Hlyabcd, \cIIIa,  \siIV, \siIIc\  and \cIIa\  absorptions are
detected  at this redshift.  Additional features  are observed  at the
expected  \wvls\ of  the associated  \nIII989 ($\lambda\ssim$2055~\AA)
and \siIIed\ ($\lambda\ssim$2470~\&~2480~\AA).   This system is at the
origin  of  the Lyman  limit  seen  at  $\sim\,$1890~\AA~ and  therefore
log~$N$(\Hly)~$>$~18.

\subsubsec{\sysabs{1.4329}}
Strong \Hly\ (from \lya\  to  \lyf) , \oVI, \siIIIa\  and \siIIc\ absorptions  are observed.
   From the  partial  Lyman limit  seen  at $\sim\,$2210~\AA,  we
derive log~$N$(\Hly)~$\sim$~17.3.

%%
%%
%% System of PC1320+4755B
%%
%%
\subsection{\qdb\ ~~\zem~=~1.11}
Only one  metal line system  is detected along  this line of  sight at
$z_{\rm  abs}$~=~0.9270. It  consists  of \Hlya,  \siIIIa, \siIV\  and
\cIV\  absorptions.  A  feature  is  observed  at  the position of  the
associated \siIIc\ ($\lambda\ssim$2430~\AA).

\section{Correlation of metal line systems} \label{sec:cormet}
%%%%%%%%%%%%
%\input metalQall.tex
%%%%%%%%%%%%
\begin{table}
\caption{List of metal lines for the four pairs}\label{tab:qmAll}
\begin{tiny}
\begin{minipage}{\hsize}
\renewcommand{\footnoterule}{}
\renewcommand{\thempfootnote}{\arabic{mpfootnote}}
\begin{tabular}{ p{25pt} p{17pt} p{5pt} p{5pt} r p{25pt} p{13pt} p{5pt} p{5pt} r}
\hline
Ident.   &$z_{\rm{abs}}$&           &$w_{\rm{obs}}$  &          &
Ident.   &$z_{\rm{abs}}$&           &$w_{\rm{obs}}$  &           \\
         &         & min$^{\rm a}$       &     max$^{\rm a}$   &   4$\sigma^{\rm b}$   & 
         &         & min$^{\rm a}$     &     max$^{\rm a}$   &   4$\sigma^{\rm b}$     \\
\hline
\multicolumn{5}{c}{LBQS0019-0145A~~(\zem=1.59)} &\multicolumn{5}{c}{LBQS0019-0145B~~(\zem=1.04)} \\\\
%\hline
           &         &       &       & 24.73  & \lya      & 0.65158 &       &  2.01 & 0.70  \\
 \cIVa:    & 0.65171 &  1.54 &  1.67 &  0.54  & \cIVa     & 0.65158 &       &  2.08 & 0.40  \\
 \alIIa    & 0.69423 &       &  0.97 &  0.69  &           &         &       &       & 0.53  \\

\\\hline
\multicolumn{5}{c}{Q0035-3518~~(\zem=1.20)} &\multicolumn{5}{c}{Q0035-3520~~(\zem=1.52)}    \\\\
%\hline
 \lya      & 1.08706 &       &  3.74 & 0.54   &           &         &       &       & 0.58  \\
 \siIIc    & 1.08706 &  0.00 &  1.56 & 0.42   &           &         &       &       & 0.61  \\
 \lya      & 1.19540 &       &  1.85 & 0.30   &           &         &       &       & 0.65  \\
 \oVIa     & 1.19540 &       &  1.12 & 0.53   &           &         &       &       & 1.33  \\
\\\hline
\multicolumn{5}{c}{Q0037-3544~~(\zem=0.84)} &\multicolumn{5}{c}{Q0037-3545~~(\zem=1.10)}    \\\\
%\hline
 \lya      & 0.59223 &       &  2.47 & 0.79   &           &         &       &       & 0.96  \\
 \cIVa     & 0.59223 &       &  0.61 & 0.56   &           &         &       &       & 0.55  \\
 \cIVa     & 0.83642 &       &  0.36 & 0.37   &           &         &       &       & 0.72  \\
\\\hline
\multicolumn{5}{c}{PC1320+4755A~~(\zem=1.56)} &\multicolumn{5}{c}{PC1320+4755B~~(\zem=1.11)}\\\\
%\hline
           &         &       &       & 0.52   & \lya      & 0.92572 &       &  4.28 & 0.78  \\
           &         &       &       & 0.79   & \cIVa     & 0.92572 &       &  2.75 & 1.05  \\
 \lya      & 1.07581 & 2.91  &  3.28 & 0.41   &           &         &       &       & 0.65  \\
 \siIIc    & 1.07581 &       &  1.07 & 0.44   &           &         &       &       & 0.58  \\
 \siIVa    & 1.07581 &       &  0.78 & 0.68   &           &         &       &       & 1.05  \\
\hline
\multicolumn{10}{l}{$^{rm a}$ minimum an maximum equivalent widths taking into account blending}\\
\multicolumn{10}{l}{$^{rm b}$ four sigma detection limit}\\
\end{tabular}
\end{minipage}
\end{tiny}
\end{table}
%%%%%%%%%%%%
We summarize in Table~6 the metal  line systems seen along one line of
sight for  which the corresponding absorption along  the adjacent line
of sight could be observed. In case of blending, $w_{\rm obs,min}$ and
$w_{\rm obs,max}$ are the minimum  and maximum equivalent width of the
transition.   These values  are computed  using  consistency arguments
relating the  equivalent widths of  lines observed in the  same system
(see examples in Sect.~5.1).   The 4$\sigma$ detection limits are also
indicated.
\par\noindent
There  are  only   6  metal  line  systems  which   have  all  $w_{\rm
r}$(Lyman-$\alpha$)~$>$~1~\AA.  Apart from the \cIV\ system at $z_{\rm
abs}$~=~0.65 toward LBQS~0019$-$0145B which may have a coincident absorption in
LBQS~0019$-$0145A  (see  Sect.~3.1.1),  none  of the  other  systems  are
detected along the adjacent line of sight down to a 4$\sigma$ limit of
$w_{\rm r}$(\lya)~$<$~0.40~\AA.  As the system at $z_{\rm abs}$~=~1.19
toward Q~0035$-$3518  could be associated with the  quasar, this means
that   out  of  5   intervening  metal   line  systems   with  $w_{\rm
r}$(Lyman-$\alpha$)~$>$~1~\AA, only one is present in the two lines of
sight.

\par\noindent
Correlation of \cIV\ systems however  has been claimed on large scales
at high redshift  (e.g. Williger et al. 1996).   Moreover, of the five
$w_{\rm  r}$~$>$~0.4~\AA~  Lyman-$\alpha$  systems  seen at  the  same
$z$~$\sim$~2 redshift in the three spectra of KP~76, 77 and 78 (triple
hits over 2-3~arcmin separations),  two show associated \cIV\ although
\cIV\   is   seen  only   in   about   one  $w_{\rm   r}$~$>$~0.4~\AA~
Lyman-$\alpha$ system out of ten  (Crotts \& Fang 1998).  All this may
indicate  that the  transverse  clustering of  \cIV\  systems is  less
pronounced at $z$~$\sim$~1 than at $z$~$\sim$~2 (see also D'Odorico et al. 2002).

\section{Correlation in the Lyman-$\alpha$ forest}\label{sec:corlya}
%%%%%%%%%%%%%%%%%%%%%
%\input lyaQallrest_th_new.tex
%%%%%%%%%%%%%%%%%%%%%
\begin{table}
\caption{List of \lya\ lines for the four pairs}\label{tab:qAll}
\begin{tiny}
\begin{tabular}{p{2pt} p{15pt} p{10pt} p{10pt} p{10pt} p{2pt} p{15pt} p{10pt} p{10pt} p{10pt} c}
\hline
    &$z_{\rm{abs}}$&$w_r^{\rm{min}}$&$w_r^{\rm{max}}$&$w_{r,4\sigma}^{\rm a}$&
    &$z_{\rm{abs}}$&$w_r^{\rm{min}}$&$w_r^{\rm{max}}$&$w_{r,4\sigma}^{\rm a}$&\\
    &         &(\AA)      &(\AA)      &(\AA)      & 
    &         &(\AA)      &(\AA)      &(\AA)      &\\
\hline
\multicolumn{5}{c}{Q0019-0145A} &\multicolumn{5}{c}{Q0019-0145B} \\
%\\%\hline
     &         &       &       &  5.64 &   5 & 0.69604 &       &  0.84 &  0.53 &\\
     &         &       &       &  1.70 &   7 & 0.79664 &       &  0.36 &  0.31 &\\
     &         &       &       &  1.50 &   8 & 0.81211 &       &  0.71 &  0.41 &\\
     &         &       &       &  0.94 &   9 & 0.84196 &       &  0.46 &  0.35 &\\
     &         &       &       &  0.91 &  10 & 0.87377 &       &  0.56 &  0.31 &\\
     &         &       &       &  0.71 &  12 & 0.90834 &  0.55 &  0.63 &  0.39 &\\
   3 & 0.91117 &  0.28 &  0.49 &  0.63 &     &         &       &       &  0.38 &\\
     &         &       &       &  0.51 &  13 & 0.94027 &       &  0.86 &  0.30 &\\
   5 & 0.94886 &       &  1.23 &  0.59 &     &         &       &       &  0.42 &\\
     &         &       &       &  0.49 &  14 & 0.98403 &       &  0.84 &  0.34 &\\
   6 & 0.98569 &       &  1.64 &  0.40 &     &         &       &       &  0.30 &\\
   8 & 1.00660 &       &  0.56 &  0.39 &     &         &       &       &  0.26 &\\
     &         &       &       &  0.30 &  15 & 1.01375 &       &  0.17 &  0.17 &\\
\\\hline
\multicolumn{5}{c}{Q0035-3518} &\multicolumn{5}{c}{Q0035-3520} \\
%\\%\hline
   1 & 0.59890 &       &  2.30 &  0.85 &     &         &       &       &  1.31 &\\
   2 & 0.66948 &  0.83 &  1.21 &  0.76 &     &         &       &       &  1.29 &\\
   5 & 0.79062 &       &  0.85 &  0.54 &     &         &       &       &  1.11 &\\
   6 & 0.81789 &       &  0.61 &  0.53 &     &         &       &       &  1.01 &\\
   9 & 0.90694 &       &  0.60 &  0.38 &     &         &       &       &  0.56 &\\
     &         &       &       &  0.57 &   1 & 0.93067 &       &  2.26 &  0.57 &\\
  10 & 1.00178 &       &  0.69 &  0.30 &     &         &       &       &  0.33 &\\
     &         &       &       &  0.42 &   7 & 1.00497 &  0.42 &  0.82 &  0.41 &\\
  11 & 1.00510 &       &  0.53 &  0.34 &     &         &       &       &  0.33 &\\
  12 & 1.03169 &       &  0.88 &  0.36 &     &         &       &       &  0.39 &\\
     &         &       &       &  0.44 &  10 & 1.04595 &  0.00 &  1.01 &  0.41 &\\
  13 & 1.08706 &       &  1.79 &  0.36 &     &         &       &       &  0.41 &\\
     &         &       &       &  0.38 &  11 & 1.08927 &  0.64 &  1.13 &  0.40 &\\
  15 & 1.16567 &  0.00 &  0.72 &  0.26 &     &         &       &       &  0.38 &\\
     &         &       &       &  0.21 &  16 & 1.16861 &       &  0.53 &  0.30 &\\
\\\hline
\multicolumn{5}{c}{Q0037-3544} &\multicolumn{5}{c}{Q0037-3545} \\
%\\%\hline
   2 & 0.59223 &       &  1.55 &  0.67 &     &         &       &       &  0.84 &\\
     &         &       &       &  0.56 &   1 & 0.71350 &       &  0.89 &  0.47 &\\
   3 & 0.72613 &       &  1.06 &  0.75 &     &         &       &       &  0.69 &\\
     &         &       &       &  0.50 &   2 & 0.76277 &       &  0.43 &  0.38 &\\
   4 & 0.77194 &       &  0.47 &  0.37 &     &         &       &       &  0.32 &\\
   5 & 0.78490 &       &  0.36 &  0.34 &     &         &       &       &  0.35 &\\
   6 & 0.80395 &       &  0.59 &  0.33 &     &         &       &       &  0.49 &\\
     &         &       &       &  0.31 &   3 & 0.82163 &       &  0.76 &  0.51 &\\
\\\hline
\multicolumn{5}{c}{PC1320+4755A} &\multicolumn{5}{c}{PC1320+4755B} \\
%\\%\hline
     &         &       &       &  1.35 &   1 & 0.62172 &       &  2.21 &  1.43 &\\
   1 & 0.62281 &  0.94 &  1.72 &  1.14 &     &         &       &       &  1.22 &\\
   2 & 0.66003 &  0.27 &  0.89 &  1.07 &     &         &       &       &  1.08 &\\
   3 & 0.66821 &  1.67 &  1.67 &  1.31 &     &         &       &       &  1.18 &\\
     &         &       &       &  1.65 &   2 & 0.75664 &       &  1.06 &  1.09 &\\
   5 & 0.76329 &  0.29 &  0.51 &  0.99 &     &         &       &       &  0.65 &\\
     &         &       &       &  1.14 &   3 & 0.77902 &       &  0.70 &  0.70 &\\
   6 & 0.84951 &       &  0.96 &  0.47 &     &         &       &       &  0.72 &\\
   7 & 0.85514 &       &  0.93 &  0.34 &     &         &       &       &  0.58 &\\
     &         &       &       &  0.32 &   4 & 0.89023 &       &  0.62 &  0.46 &\\
     &         &       &       &  0.48 &   6 & 0.92572 &       &  2.22 &  0.72 &\\
  11 & 0.92713 &  0.97 &  1.04 &  0.40 &     &         &       &       &  0.61 &\\
  14 & 0.95579 &       &  1.78 &  0.43 &     &         &       &       &  0.80 &\\
  15 & 0.97471 &       &  0.37 &  0.29 &     &         &       &       &  0.54 &\\
     &         &       &       &  0.36 &   7 & 1.02501 &       &  0.64 &  0.51 &\\
  19 & 1.07581 &  1.40 &  1.58 &  0.22 &     &         &       &       &  0.41 &\\
     &         &       &       &  0.33 &   9 & 1.08477 &       &  0.71 &  0.38 &\\
\hline
\multicolumn{10}{l}{$^{\rm a}$ calculated using the width of the line detected along}\\
\multicolumn{10}{l}{either line of sight.}\\
\end{tabular}
\end{tiny}
\end{table}
%%%%%%%%%%%%%%%%%%%%%
\subsection{The \lya\ line list}
From  the line lists  obtained as  described in  Sect.~2 and  given in
Tables 2 to 5, we have  extracted for each pair of QSOs, a master line-list of
Lyman-$\alpha$  lines based on  several criteria:  (i) we  include all
isolated lines when  no other identification is found;  (ii) the lines
must  be at  more  than  3000~km~s$^{-1}$ blueward  of  the two  \lya\
emission lines;  (iii) we use physical consistency  arguments to infer
the  presence of Lyman-$\alpha$  lines blended  with metal  lines (for
example  \cIVa\ cannot  be  weaker  than \cIVb);  only  limits on  the
equivalent  width  can be  inferred  this  way;  (iv) we  impose  some
equivalent   width  threshold.   The   Lyman-$\alpha$  line   list  is
summarized in  Table~7. The  columns correspond to:  \#1 and  \#6 line
number  in  the spectrum  (see  Fig.~\ref{fig:spectra});  \#2 and  \#7
Lyman-$\alpha$ redshift;  \#3:\#4 and \#8:\#9 the  range in equivalent
widths in  case of  blending (see below);  \#5 and \#10  the 4$\sigma$
equivalent width detection limit using  the width of the line detected
along either line of sight.
\par\noindent
In  \qaa, absorption  \#3  cannot only  be  \oVIb\ as  it has  $w_{\rm
obs}$~=~2.10~\AA~  whereas   $w_{\rm  obs}$(OVI1031)~=~1.64~\AA.   The
hidden  Lyman-$\alpha$ line  has  0.28~$<$~$w_{\rm obs}$~$<$~0.49~\AA~
corresponding to the \oVI\ doublet ratio ranging from 1 to 2.
\par\noindent
Absorption \#8  coincides in  redshift with \cIII\  at \zabs{1.49672}.
However, we  consider this identification unlikely.   Indeed, the line
is quite  strong ($w_{\rm obs}$~=~1.12~\AA)  even though the  UVES and
STIS spectra do not show any other metal lines at this redshift except
for a very weak \mgIIb\  line system and a 2.5$\sigma$ feature shifted
by  1.5~\AA\ from  the expected  position of  \siIIIa.   Therefore, we
identify this line as Lyman-$\alpha$ at \zabs{1.00660}.
\par\noindent
In \qab, the limits on the equivalent width of the Lyman-$\alpha$ line
blended with  \siIVb\ at $z_{\rm  abs}$~=~0.65381 in feature  \#12 are
derived applying the doublet ratio to the \siIVa\ equivalent width.
\par\noindent
In \qba, we estimate the \ew\  of \lyc\ (line \#2) from the associated
Lyman-$\alpha$ and \lyb\ absorptions.
\par\noindent
In  \qbb,  for   feature  \#7, we  assume  that   \cIIIa\  at  $z_{\rm
abs}$~=~1.49469  contributes very  little since  the  associated \lya\
absorption is weak and no other  metal line is detected; limits on the
\ew\ of  the Lyman-$\alpha$ line at $z_{\rm  abs}$~=~1.00497 come from
limits  on \lyc\  at $z_{\rm  abs}$~=~1.50621 derived  from  the \lya\
absorption.   Similarly, we  have  constrained the  \ew\  of \lyb\  at
$z_{\rm  abs}$~=~1.47617  in  the  absorption feature  \#11  from  the
associated \lya\ line.
\par\noindent
In  \qda, the  limits on  the \ew\  of \lyd\  (feature \#1)  and \lyc\
(feature \#2)  at $z_{\rm  abs}$~=~1.07575 come from  their associated
\lyb\ and  Lyman-$\alpha$ absorptions. For feature \#3  we assumed for
the  \ew\  of  \cIIIa\  a  conservative limit  of  1.5~\AA\  from  the
associated \cIIa\ line. Finally, lower and upper limits on the \ew\ of
\oVIa\ at $z_{\rm abs}$~=~1.07726 (feature \#5) and \feIIe\ at $z_{\rm
abs}$~=~0  (feature \#11)  come from  \oVIb\ which  has  an equivalent
width $w_{\rm obs}\ssim0.8$~\AA,  below the 4$\sigma$ detection limit,
and \feIIa\ respectively.
\par\noindent
Note that  the number of metal  lines which could  be misidentified as
Lyman-$\alpha$  lines is  expected to  be small.  Indeed, they  can be
neither  Mg~{\sc  ii} nor  Fe~{\sc  ii}  lines  because the  strongest
potential  lines of these species  are  redshifted   beyond  the  wavelength   of  the
Lyman-$\alpha$  line with the  highest redshift  in our  sample.  They
cannot  be lines  too close  in  wavelength to  Lyman-$\alpha$ as  our
careful procedure would have identified them.  The only possibility is
that some isolated Al~{\sc ii}$\lambda$1670 or C~{\sc iv}$\lambda$1548
(with C~{\sc  iv}$\lambda$1550 not detectable) lines  could be present
in the wavelength range where  we search for coincidences.  The number
of C~{\sc iv} and Mg~{\sc ii} systems with $w_{\rm r}$~$>$~0.3~\AA~ at
$z$~$\sim$~0.3  is  0.87  and  0.75  per  unit  redshift  respectively
(Bergeron et al. 1994, Boiss\'e et al. 1992). We consider that half of
the Mg~{\sc ii}  systems have an associated Al~{\sc  ii} line and half
of  the C~{\sc  iv} systems  would have  only  C~{\sc iv}$\lambda$1548
detected.  Moreover, only less  than two-third  of these  systems have
$w_{\rm r}$~$>$~0.4~\AA~  which is the  usual 4~$\sigma$ limit  of our
spectra.   As  our  survey  samples  a redshift  interval  of  $\Delta
z$~=~2.96 (considering six lines of sights, see Table~1), the expected
number of misidentified lines is of the order of 1 to 2. This is to be
compared to the 46  Lyman-$\alpha$ lines with $w_{\rm r}$~$>$~0.4~\AA~
we detect. Note  that the latter number is  consistent with the number
of lines detected in the {\sl HST} Key-program (Jannuzi et al. 1998).
\subsection{Correlation}
\figcor
\figcorI
From  the   master  line  list,   we  selected  lines   with  $w_{\rm
r}>$\nolinebreak0.3~\AA\  and applied  the Nearest-Neighbor  method as
described  in  e.g.  \cite{You01}  (2001)  to estimate  the  level  of
correlation  between  absorptions  detected  along adjacent  lines  of
sight.  In this method, there is no a priori velocity separation limit
in  the definition  of  a coincidence.  A  couple of  lines along  two
different \loss\ is declared to be  a coincidence if each of the lines
is  the  nearest  neighbor  of  the other.  Note  that  the  procedure
underestimate the clustering  signal as it does not  take into account
the difference of S/N ratio along the two lines of sight.
\par\noindent
The distributions of velocity  separations (\adv) between the two lines
in a coincidence are plotted in Fig.~\ref{fig:corr} for the pairs with
angular separation \ssim2~arcmin (Panel a), 3~arcmin (Panel b) and the
complete sample  (Panel c).  To  improve statistics, we have  added to
our results, the  data of \cite{You01} (2001) on  two additional pairs
separated by \ssim2~arcmin and  \ssim3~arcmin. The results are plotted
in Fig.~\ref{fig:corr} for the pairs  with an angular  separation \ssim2
(Panel d), 3~arcmin (Panel e) and for all the pairs (Panel f).
\par\noindent
To estimate the excess of  correlation with respect to randomly placed
absorption lines, we produced 100000 simulated master line lists drawn
from a population of randomly redshifted lines, taking the same number
of lines and the same \wvl\  range as in the observed spectra. Results
of applying the  same method to the simulated line  lists are given as
dotted  lines in  Fig.~\ref{fig:corr}. The  error bars  in  the figure
correspond to the  rms of the values found in  the simulation.  As the
corresponding  distribution  is  not  Gaussian,  we  indicate  in  the
following  the probability  that the  observed number  of coincidences
occurs in the simulated population.
\par\noindent
We detect 2 and 4 coincidences with \adv\ smaller than 500~km~s$^{-1}$
in  the  two  pairs   separated  by  \ssim2~arcmin  and  \ssim3~arcmin
respectively. The fact  that the number of coincidences  is smaller in
the closest pairs is  not statistically significant. When the complete
sample is  used (Panel  c of Fig.~\ref{fig:corr}),  the excess  in the
first bin  (\adv$<$\nolinebreak500~km~s$^{-1}$) is significant  at the
99.20\%  level.  To  increase the  statistics,  we have  added to  our
sample data from \cite{You01} (2001). The total number of coincidences
with  \adv$<$\nolinebreak500~km~s$^{-1}$ and  $w_{\rm r}$~$>$~0.3~\AA~
is increased to 6, 8 and  14 for the pairs separated by \ssim2~arcmin,
\ssim3~arcmin  and  the complete  sample.  The corresponding  excesses
relative to  simulations of randomly  placed lines are  significant at
the 95.57\%,  99.92\% and 99.97\%  levels respectively. The  excess is
about the  same when no threshold  is applied to  the equivalent width
(see Fig.~\ref{fig:corrI}).  In that case, the  number of coincidences
in the  first bin is  20 and the  excess is detected at  the 4$\sigma$
level.
\figcorIbin
\par\noindent
This  clearly shows that  the Lyman-$\alpha$  forest is  correlated on
scales larger  than 1$\,h^{-1}_{50}$~Mpc proper at  $z$~$\sim$~1. However, we
should note  that, in the complete  sample, 12 of  the 20 coincidences
with \adv\  smaller than  500~km~s$^{-1}$ actually have  \adv\ greater
than  200~km~s$^{-1}$.  This  is  also  the  case  for  8  of  the  14
coincidences  in the $w_{\rm  r}$~$>$~0.3~\AA\ sample.   These velocity
differences are  probably related to peculiar  velocities of different
objects and reveal that the scale  we probe is not related to the real
size of individual absorbers.
\par\noindent
\figw1vsw2
\par\noindent
In Fig.~\ref{fig:w1vsw2}  we plot  the equivalent width  observed along
one  line of  sight versus  the  equivalent width  observed along  the
second  line of  sight  for  the 20  pairs  with velocity  differences
smaller  than  500~km~s$^{-1}$.   There  is  no  clear  trend  in  the
plot. This is not really surprizing as we do not expect the absorption
strengths to be correlated at such a separation neither in the case of
a common absorber nor in the case of independent objects.
\section{Conclusion}
We  have searched for  coincidences of  \lya\ absorbers  at $z~\leq$~1
along the lines of sight  toward four pairs of \qsos\ with separations
2$-$3~arcmin,  observed with \HST\  STIS.  Using  the Nearest-Neighbor
statistics,  we  have constructed  the  distribution  of the  velocity
difference between absorption lines  detected along two adjacent lines
of sight.  We have compared this observed distribution to that derived
from  Monte-Carlo simulations  placing at  random the  same  number of
lines in the same wavelength ranges as in the observations.  For lines
with $w_{\rm r}$~$>$~0.3~\AA~, we  find an excess of coincidences with
velocity separations  smaller than 500~km~s$^{-1}$  significant at the
99.2\% level. Combining  our data with those in  the literature (Young
et al. 2001), the excess relative to simulations is significant at the
99.97\%  level for  lines  with $w_{\rm  r}$~$>$~0.3~\AA~  and at  the
99.98\% level if no condition on the equivalent width is imposed.
\par\noindent
The  result is  consistent with  the  findings of  similar studies  at
higher  redshift  (\cite{Cro98}  1998,  \cite{Wil00} 2000).  There  is
however an important difference between high and intermediate redshift
observations.   At high  redshift,  the excess  is  seen for  velocity
separations    between    two    coincident   lines    smaller    than
200~km~s$^{-1}$. At  lower redshift,  the mean velocity  difference is
larger (see Fig.~\ref{fig:corr1bin}).  This is not a systematic effect
related  to the  low  spectral resolution  of  our data.  It could  be
related  to  the  increase  of  peculiar  velocities  with  decreasing
redshift   for  comparable   spatial  scales.    This   conclusion  is
strengthened  by  the  fact  that the  transverse  correlation,  $\chi
\propto    \sum_i{(1-F_1(\lambda_i))\times(1-F_2(\lambda_i))}$   where
$F_1(\lambda_i)$  and  $F_2(\lambda_i)$  are  the two  QSO  normalized
fluxes  at wavelength  $\lambda_i$,  at $z$~$\sim$~1  measured on  our
spectra is  very small, $|\chi|<0.01$, whereas  it is of  the order of
0.2  at $z$~$\sim$~2  for  the  same separation  (Rollinde  et al.  in
preparation, see also McDonald 2000, 2001 and Viel et al. 2001).  This
can be  explained by the fact that  large values of $\chi$  are due to
coincidences  with  velocity  separations  smaller than  the  spectral
resolution. In conclusion, evolution from  high to low redshift is not
seen in the level of correlation but rather in the velocity difference
between  lines  of sight  which  increases  with decreasing  redshift.
Simulations  have shown  that  absorption lines  with  a given  column
density correspond to higher overdensities at low-redshift compared to
high redshift and most of the Lyman-$\alpha$ forest at low redshift is
to  be revealed by  very weak  lines (\cite{Rie98}  1998; \cite{The98}
1998;  Penton  et  al.   2000).   Therefore,  detailed  study  of  the
cosmological  evolution  of the  Lyman-$\alpha$  forest  will only  be
possible when the  sensitivity of the instruments will  be high enough
to routinely detect lines with $w_{\rm r}$~$<$~0.01~\AA\ in the UV.

\begin{acknowledgement}
{We thank  C\'edric Ledoux for the reduction  of the LBQS~0019$-$0145A
UVES spectrum.  This work  was supported in  part by the  European RTN
program  ``The Intergalactic  Medium''  and by  a  PROCOPE program  of
bilateral collaboration between France  and Germany.  BA and PPJ thank
AIP for hospitality. }
\end{acknowledgement}

%======================================

\appendix

\listofobjects
\end{document}